\documentclass{aastex}
\usepackage{emulateapj5}

\def\mathnew{\mathsurround=0pt}
\def\jetp{Sov.~Phys.-JETP}
\def\simov#1#2{\lower .5pt\vbox{\baselineskip0pt \lineskip-.5pt
\ialign{$\mathnew#1\hfil##\hfil$\crcr#2\crcr\sim\crcr}}}

\def\simgreat{\mathrel{\mathpalette\simov >}}

\shorttitle{Model for QPO soft lags}
\shortauthors{Lee, Misra \& Taam}

\begin{document}

\title{A Compton Up-scattering Model for Soft Lags in the Lower Kilohertz QPO in
4U1608$-$52}

\author{Hyong C. Lee\altaffilmark{1}, R. Misra\altaffilmark{2} and Ronald
E. Taam\altaffilmark{3}} 
\vspace*{0.3cm}
\affil{Department of Physics and Astronomy, Northwestern University,\\
       2145 Sheridan Road, Evanston, IL 60208, USA.}
\altaffiltext{1}{E-mail: hyongel@northwestern.edu}
\altaffiltext{2}{E-mail: ranjeev@finesse.astro.northwestern.edu}
\altaffiltext{3}{E-mail: taam@apollo.astro.nwu.edu}

\begin{abstract}

An empirical Compton up-scattering model is described which reproduces both 
the fractional amplitude (RMS) vs. energy and the soft time lags in 
the $\approx$ 830 Hz QPO observed in 4U1608-52 on Mar. 3, 1996. A combination 
of two coherent variations in the coronal and soft photon temperatures 
(with their relative contributions determined by enforcing energy conservation) 
gives rise to the QPO's energy dependent characteristics.
All input parameters to the model, save a characteristic plasma size
and the fraction of Comptonized photons impinging on the soft photon source,
are derived from the time-averaged photon 
energy spectrum of the same observation.  Fits to the fractional RMS and phase 
lag data for this kilohertz QPO imply that the 
spatial extent of the plasma is in the range from $\sim$ 4 to 15 km. 
\end{abstract}


\section{Introduction}

Soon after the launch of the Rossi X-ray Timing Explorer (RXTE),
rapid (300--1300~Hz), nearly periodic variability in the X-ray light curves 
of low mass X-ray binary systems (LMXB's) was discovered (Strohmayer et al.~1996, 
van~der~Klis et al.~1996).  These oscillations, referred to as kilohertz QPO 
(quasi-periodic oscillations), have now been observed in over two dozen neutron 
star bearing LMXB's (see van~der~Klis~2000 for a review).  They are distinguished 
by high frequencies and high quality factors (Q=FWHM/frequency), and tend to 
be seen in pairs, with nearly constant frequency separation between the lower and 
higher frequency peaks (also called the lower and upper QPO). 
 The high frequencies of kilohertz QPO are thought to tie them to phenomena 
taking place in the inner regions of accretion disks surrounding neutron stars,
making them potentially valuable probes of strong gravity.

A broad range of theoretical ideas have been proposed to explain the phenomenology 
of kilohertz QPO.  In these models, one of the frequencies is generally identified 
as the Keplerian frequency of the innermost orbit of an accretion disk.  The sonic point
model (Miller et al.~1998) identifies the second frequency as 
the beat of the primary QPO with the spin of the neutron star.  Stella \& Vietri 
(1999) have proposed a general relativistic precession/apsidal motion model wherein 
the primary frequency is the Keplerian frequency of a slightly eccentric
orbit, and the secondary is due to the relativistic apsidal motion of this
orbit.  On the other 
hand, in the two oscillator model (Osherovich \& Titarchuk~1999, 
applied to 4U1608$-$52; Titarchuk \& Osherovich 1999), the secondary 
frequency is due to the transformation of the primary (Keplerian) frequency in the 
rotating frame of the neutron star magnetosphere.  The strength (and weakness) of these
models lie in their ability (or inability) to predict the variation of the frequency 
separation with the kilohertz QPO frequency and the variation of their frequency with 
other low frequency QPO observed in the source (see van der Klis 2000 for a comparison 
of recent observations with model predictions).  However, these 
dynamic models generally do not address how these dynamic oscillations
affect, or couple to, the radiative processes which finally give rise 
to the oscillations seen in the X-ray spectra.  The observed energy dependent 
features of the QPO, namely the strength of the QPO vs. photon energy 
(fractional root mean square, or RMS, vs. energy) and the phase lag of fluctuations at
different photon energies relative to variations at a reference energy, are expected to
depend primarily on the radiative mechanism and its coupling to the dynamic
behavior of the system.

While a unified model for kilohertz QPO, which self-consistently incorporates both the dynamical 
and radiative mechanisms, remains illusive, significant progress can be achieved by 
studying the radiative response of a system to an oscillating perturbation.  Such an 
analysis would make definite predictions about the energy dependent features of
the QPO, thereby placing constraints on the radiative mechanism, and perhaps, shed light on the nature
of the coupling between the dynamical and radiative processes.
 
The power law photon energy spectra often observed in LMXB's suggests that the 
dominant radiative mechanism in these systems is Compton up-scattering of 
soft photons in a hot plasma (Sunyaev \& Titartchuk~1980, Podznyakov et
al.~1983).  The temporal behavior of such models has been studied in a variety of 
circumstances (e.g.  Wijers et al.~1987; Hua, Kazanas \& Titarchuk~1997; 
Bottcher \& Liang~1999).  For QPO in particular, both the fractional amplitude vs. 
energy and the relative time lag vs. energy can be predicted based upon models where
a hot, uniform plasma is illuminated by soft black body photons, accounting for 
the system's response to variations in either the soft photons or the plasma 
(Wijers 1987, Lee \& Miller~1998). 
 
In Compton up-scattering models, a photon's mean residence time in the hot
plasma increases with the energy of the escaping photon since, on average, 
higher energies require a larger number of scatters.  Thus it would seem 
that for these models, variability in high energy photons should be delayed 
relative to their lower energy counterparts (i.e., a hard time lag is expected).
 In fact, analysis of simple Comptonization models, where the variability is
caused only by oscillations in the intensity of the soft input spectrum, reveal
that the time lag between the high energy, $E_h$, and low energy, $E_l$, photons
should roughly scale as
log$(E_h/E_l)$ (Nowak \& Vaughan~1996). 
 It is therefore rather surprising that soft lags
(i.e., variability in soft photons delayed with respect to hard photons) 
seem to be the norm in kilohertz QPO (Vaughan et al.~1998, Kaaret et al.~1999,
Markwardt et al.~1999).  This inconsistency has led to the suggestion that Compton down-scattering 
could be the dominant radiative process in these systems.  However, it is difficult 
to reconcile the observed increase in the amplitude of the oscillations with energy 
with such models.  Moreover, this would point to the unlikely 
scenario in which the radiative process for the time-averaged spectrum (Compton 
up-scattering) differs from that for the QPO oscillations (Compton 
down-scattering).

This paper describes a Compton up-scattering model that gives a consistent 
interpretation of the time-averaged photon spectra as well as the energy
dependence of both the fractional RMS amplitude and time lag for the 830~Hz QPO
detected in 4U1608$-$52 on March 3, 1996 (Berger et al.~1996).  The
observed soft lag is a natural consequence of the model provided that 
the temperatures of the plasma and of the soft photon source
oscillate coherently at the QPO frequency.  It is further shown that
coherent oscillation of the temperatures is expected when energy transfer
between the plasma and the soft photon source is taken into account. The model
imposes limits on the size of the Comptonizing region and places constraints on
any future model invoked to explain how the dynamical origin of the
QPOs couples with the Comptonizing medium in LMXB's. 

\section{The Compton up-scattering model}
	
Consider a uniform hot plasma characterized by its temperature ($T_c$), 
size ($l$) and optical depth $\tau = n_e \sigma_T l$, where $n_e$ is the electron 
number density and $\sigma_T$ is the Thomson cross section.  The hot plasma 
up-scatters soft photons from an external source.  Unlike previous work, the soft
photons are assumed to be described by a Wien spectrum (at a temperature $T_s$) 
instead of a black body.  Thus the soft photon source lies in a  
Comptonizing region with an optical depth large enough that the emergent 
spectrum saturates to a Wien peak.  The choice of a Wien spectrum rather than a black 
body spectrum is motivated by the expectation that (for $T_s \approx 1$ keV and for reasonable 
electron densities $ << 10^{25}$ cm$^{-3}$), the photon density inside the source 
will not be in thermal equilibrium.  It should be noted that broad band 
observations of the black hole system Cygnus X-1 reveal that the soft photon 
component (which is assumed to provide the seed photons for Comptonization) is better 
described as a Wien peak rather than a black body or a sum of black bodies
(Di Salvo et al.~2000).  Since the soft photon source could itself be heated by 
the Comptonized photons from the hot plasma, it is assumed that a fraction, 
$f$, of the Comptonized photons is incident on the soft photon source.  In the 
framework of this model, oscillations in the photon escape rate from the hot 
plasma are due to corresponding coherent oscillations in the plasma temperature 
($T_c$) and the soft photon source temperature ($T_s$) while the size ($l$), 
the electron density ($n_e$) and the total rate of photons emerging from the 
plasma remain constant. 

For non-relativistic temperatures ($kT_c << m_e c^2$) and low photon energies 
($E << m_c c^2$), the evolution of the photon density inside the hot plasma  
is governed by the simplified Kompaneet's equation (with the induced scattering 
term neglected) (Kompaneets 1957)
\begin{eqnarray}
t_{c} {d n_\gamma \over d t } &  = &{1\over m_ec^2}{d \over d E}[-4kT_c E n_\gamma  + E^2n_\gamma + kT_c {d \over d E} (E^2 n_\gamma)] \nonumber\\
& & + t_c\dot{n}_s - t_c \dot{n}_{esc}\,.
\label{eq:komp}
\end{eqnarray}
Here $n_\gamma$ is the photon density inside the plasma, $E$ is the energy, and 
$t_c = l/c \tau$ is the Thomson collision time scale.  The number of photons per 
second per unit volume injected into the plasma from the soft photon source is
$\dot n_{s} = C T_s^{-3} E^2 exp(-E/kT_s)$ where $C$ is a normalization constant. 
 This is balanced by the corresponding rate of photons escaping from the plasma 
$\dot n_{esc} \approx n_\gamma/(N_{esc} t_c)$ where $N_{esc} \approx \tau 
(\tau +1)$ is the average number of scatterings a photon undergoes before escaping. 

The time-averaged spectrum arising from the plasma can be estimated
using $\dot {n}_{esc0} \approx n_{\gamma 0}/(N_{esc} t_c)$, where
$n_{\gamma 0}$ is the steady state photon number density corresponding to
time-averaged values of plasma temperature ($T_{c0}$) and soft
photon Wien temperature ($T_{s0}$).  For $kT_{s0} << E << kT_{c0}$,
it can be shown from equation (\ref{eq:komp}) that the
spectrum is a power-law with photon index ($\Gamma$) related to
$N_{esc}$ by
\begin{equation}\label{eq:nescape}
N_{esc}=\frac{(m_{e}c^2/kT_c)}{\Gamma^2+\Gamma - 2}.
\end{equation}

Lee \& Miller (1998) studied the temporal behavior of a Comptonizing
medium by computing the linear response of the photon spectrum to a sinusoidally
varying physical parameter at a given angular frequency $\omega$.  Following
their formulation, the plasma temperature is
approximated as $T_{c} = T_{c0}(1 + \delta T_c e^{-i\omega t})$ and
the soft photon temperature as  $ T_s = T_{s0}(1 + \delta T_s e^{-i\omega t})$
where $\delta T_c$ and $\delta T_s$ are, in general, small complex quantities.
The resulting time varying photon number density is taken to be
$n_\gamma = n_{\gamma 0}(1 + \delta n_{\gamma} e^{-i\omega t})$ with the
output spectrum given by $\dot n_{esc} = n_{\gamma}/(N_{esc} t_c)$.
 $\delta n_\gamma$ is related to $\delta T_{c}$ and $\delta T_{s}$ 
by the linearized Kompaneets equation (Lee \& Miller~1998):
\begin{eqnarray}
\label{eq:linkomp}
( \frac{1}{N_{\rm esc}} - i\omega t_{\rm c}) n_{\gamma 0} \delta n_\gamma &  = & { - 1\over m_ec^2}\,{d\,\over dE}\big [4E kT_{c0} n_{\gamma 0} ( \delta T_c + \delta n_\gamma )  \nonumber \\
& & - E^2 n_{\gamma 0} \delta n_\gamma \nonumber \\
& & - {d\,\over dE} E^2 kT_{c0} n_{\gamma 0} ( \delta T_c + \delta n_\gamma )\big ] \nonumber \\
& & -t_c \dot {n}_{s0} ({E \over k T_{s0}} - 3) \delta T_s\; .
\end{eqnarray}
The above equation implicitly assumes that
the soft photon source spectrum $\dot n_{s}$ is a Wien peak with
a constant photon number flux.  It is useful to define $\delta n_{\gamma c}$ as 
the photon variation $\delta n_\gamma$ corresponding to a unit variation
in the plasma temperature ($\delta T_c = 1$) and similarly
$\delta n_{\gamma s}$ corresponding to a unit variation in the soft
photon temperature ($\delta T_s = 1$).  Thus, $\delta n_{\gamma c}$
($\delta n_{\gamma s}$) can be computed from equation (\ref{eq:linkomp}) 
by setting
$\delta T_c = 1$ ($\delta T_s = 1$) and $\delta T_s = 0$ ($\delta T_c = 0$).
Since only the linear response is being considered,
the variation in total photon number density can be written as
\begin{equation}
\label{eq:tot_var}
\delta n_\gamma = \delta T_c \delta n_{\gamma c} + \delta T_s \delta n_{\gamma s} \; .
\end{equation}
Extending this method to include higher order terms leads to harmonics of
the fundamental represented by the linear response.  In the
absence of any observed kilohertz QPO harmonics and in view of the simple assumptions
made in this model, only the linear response will be considered.

Variations in the plasma temperature will induce a variation in the plasma
luminosity.  Since a fraction ($f$) of the Comptonized photons re-enter
and heat up the soft photon source, a variation in the plasma luminosity 
causes a variation in the
soft photon source temperature.  Thus $\delta T_c$ and $\delta T_s$ are
related by
\begin{equation}
\label{eq:Ts1}
\delta T_s \approx f A {\delta L_c \over L_c}  = f A {\int n_{\gamma 0} \delta n_{\gamma} (E) E dE \over \int n_{\gamma 0}  (E) E dE}
\end{equation}
where $A$, the Compton amplification factor, is defined as the ratio
of the plasma to soft photon source luminosities.  Using equation
(\ref{eq:tot_var}), the above equation can be written as
\begin{equation}
\label{eq:Ts2}
\delta T_s \approx \delta T_c   { f A \int n_{\gamma 0} \delta n_{\gamma c} E dE
\over   \int n_{\gamma 0} (1 - f A\delta n_{\gamma s}) E dE } .
\end{equation}

For a given variation in plasma temperature ($\delta T_c = |\delta T_c|$) and
time-averaged model parameters ($T_c$, $T_s$, $N_{esc}$, $l$, $A$ and $f$),
the fluctuations in the photon number density ($\delta n_\gamma$) can be obtained
from equations (\ref{eq:linkomp}), (\ref{eq:tot_var}) and (\ref{eq:Ts2}).
The complex quantity $\delta n_\gamma (E) $ leads to  
the time lag vs. energy and amplitude vs. energy for the QPO.  Since
the disk parameters ($T_c$, $T_s$, $N_{esc}$ and $A$) can, in principle,
be constrained by fitting the time-averaged spectrum of the source,
$\delta n_\gamma (E)$ depends only on
the amplitude of the plasma temperature variation, the
plasma size, and the fraction of photons incident on the soft photon source.
	
\section{Results}

The predictions of the Comptonization model described in \S 2 are 
compared with the RXTE (PCA) observations of the LMXB 4U1608$-$52 on 
March 3, 1996 (Berger et al. 1996), where a 830~Hz QPO (the lower peak --
see Mendez et al.~1997) was discovered.  The time-averaged spectrum from
this observation is well described by the model with a reduced 
$\chi^2 = 1.03$ in the energy range 3 - 20 keV.  The soft photon Wien
temperature and the photon index are well constrained to be  
$k T_s = 1.11 \pm 0.04$ keV and $\Gamma = 3.55 \pm 0.05$ respectively.
 However, since no cutoff in the spectrum was observed, spectral fitting
imposed only a lower limit on the plasma temperature $k T_c > 20$ keV.  Since the
Kompaneet's equation (\ref{eq:komp}) is valid only for $k T_c << m_e c^2$,
$kT_c = 25$ keV has been chosen for the analysis (choosing $k T_c = 50$ keV 
makes no significant qualitative difference in the results). 
These values of $k T_c $ and $\Gamma$ give $N_{esc} = 1.44$ (equation
\ref{eq:nescape}) and a Compton amplification factor $A \approx 1.33$.

The model predictions and RXTE observations of the amplitude 
(root mean square, or RMS, fraction) as a function of energy for the 830 Hz QPO 
(Berger et al.~1996) are compared in Figure 1, for 
different values of $l$ and $f$. The overall 
normalization of the curve is fixed by choosing a best fit value of $|\delta T_c|$. 
The predicted amplitude is zero at $E \approx 4$ keV (comparable to the peak of 
the input spectrum) since the emergent spectrum pivots about this energy. This  
behavior is expected as oscillations in the soft photon Wien temperature and in 
the plasma temperature both preserve photon number.  The lag is illustrated   
as a function of energy in Figure 2 (Vaughan et al.~1998) compared to the model 
prediction for the same values of $l$ and $f$ used in Figure 1. Reasonable fits 
to both variations are obtained for $4$ km $ < l < 15$ km and $0.5 < f < 0.85$.

Since the amplitudes of plasma and Wien temperature oscillations are nearly equal
(i.e. $|\delta T_c| \approx |\delta T_s|$), the photon fluctuations at high
energies, $\delta n_{\gamma H} = \delta n_\gamma (E \approx 15$ keV), are predominately
caused by $\delta T_c$ while the photon fluctuations at low
energy, $\delta n_{\gamma L} =\delta n_\gamma (E \approx 3$ keV), are predominately caused  by $\delta T_s$. 
Thus, the time lag between $\delta T_c$ and $\delta n_{\gamma H}$ is 
$\Delta t_H \approx N_H t_c$, where $N_H \approx 9$ is the average number of scatters a high energy photon undergoes.
The time lag between $\delta T_c$ and $\delta n_{\gamma L}$ can be
expressed as $\Delta t_L \approx  t_c + \Delta t_{sc}$, where $\Delta t_{sc}$ is the time lag between ($\delta T_c$) and
($\delta T_s$). Thus the time lag between high and low energy photons is $\Delta t_{hs} \approx 8 t_c - \Delta t_{sc}$,
so that when $\Delta t_{sc} > 8 t_c$, the system will show soft lags.


If the soft photon source is assumed to be a black body instead
of a Wien peak, an additional ad hoc oscillation in the soft 
photon number flux is required for the predicted RMS fraction vs. energy to agree
with observations.  Thus, a Wien spectrum is favored for the soft photons,
since it minimizes the number of oscillating components needed.

In our model, the primary oscillation takes place in the plasma with the 
soft photon source responding to the variation. Alternately, it is possible
that the soft photon Wien temperature oscillates and, in order to keep the total
luminosity of the plasma constant, induces variability in the plasma
temperature.  However, contrary to observations, such a model does 
not predict soft lags.  Thus it seems that the QPO manifests itself only as 
an oscillation in the plasma temperature, rather than variations in other
parameters of the Comptonization model.

The large value for the fraction of Comptonized photons incident on  
the soft photon source, $f \simgreat 0.5$, suggests that the geometry
of the hot plasma can be described as a corona on top of
a cold accretion disk.  The size of the corona ($l \approx 5$ km) indicates
that it is a significant
fraction of the radius of the disk.  This result implies that the nearly
coherent variability making up the kilohertz QPO is not limited to a narrow
region and hints at the existence of a global mode in 
the corona.  In such an interpretation, the high Q values ($\sim 100$) 
characteristic of the high frequency oscillations would provide a strong constraint 
on the excitation and damping of the mode (Nowak et al.~1997).
We point out that the Comptonizing model does not provide insight into the 
the detailed geometry of the source.  An alternative geometry would be an X-ray 
source (i.e., the neutron star) surrounded by a flaring accretion disk. This
geometry introduces an additional time lag due to the light 
travel time between the Comptonizing plasma and soft photon source.

Since the energy spectra did not exhibit an observable cutoff,
the plasma temperature could be significantly larger than the assumed $25$ keV. 
 In that case, the average energy transfer per scatter would become a significant
fraction of the photon energy and the Kompaneet's equation, and the model
predictions with it, would no longer be
accurate (Katz 1987).  Hence, simple extensions of our model to higher plasma
temperatures are not useful, and detailed 
fitting to the observations will be warranted only when models incorporating
non-standard geometries in sophisticated time-dependent Monte Carlo methods 
(incorporating energy balance) are  
developed.  In this light, the ability of the simple model presented here to
explain the main features of a kilohertz QPO's energy dependent properties is
highly encouraging.

In the future, this work will be extended to other systems exhibiting QPOs,
particularly to the upper kilohertz QPO of the doublet observed in neutron star systems. 
Such an endeavor will not only check the robustness of the results presented 
here, but will also provide further constraints on theoretical models of the QPO 
phenomenon.

\acknowledgements
HL acknowledges the support of NASA S98-GSRP-129.  RM acknowledges the support
of the Lindheimer Fellowship.  This research has made use of data obtained
through the High Energy Astrophysics Science Archive Research Center
at the NASA/Goddard Space Flight Center.

\input psbox.tex
\begin{figure*}[h]
\hspace{-1.5cm}
{\mbox{\psboxto(17cm;20cm){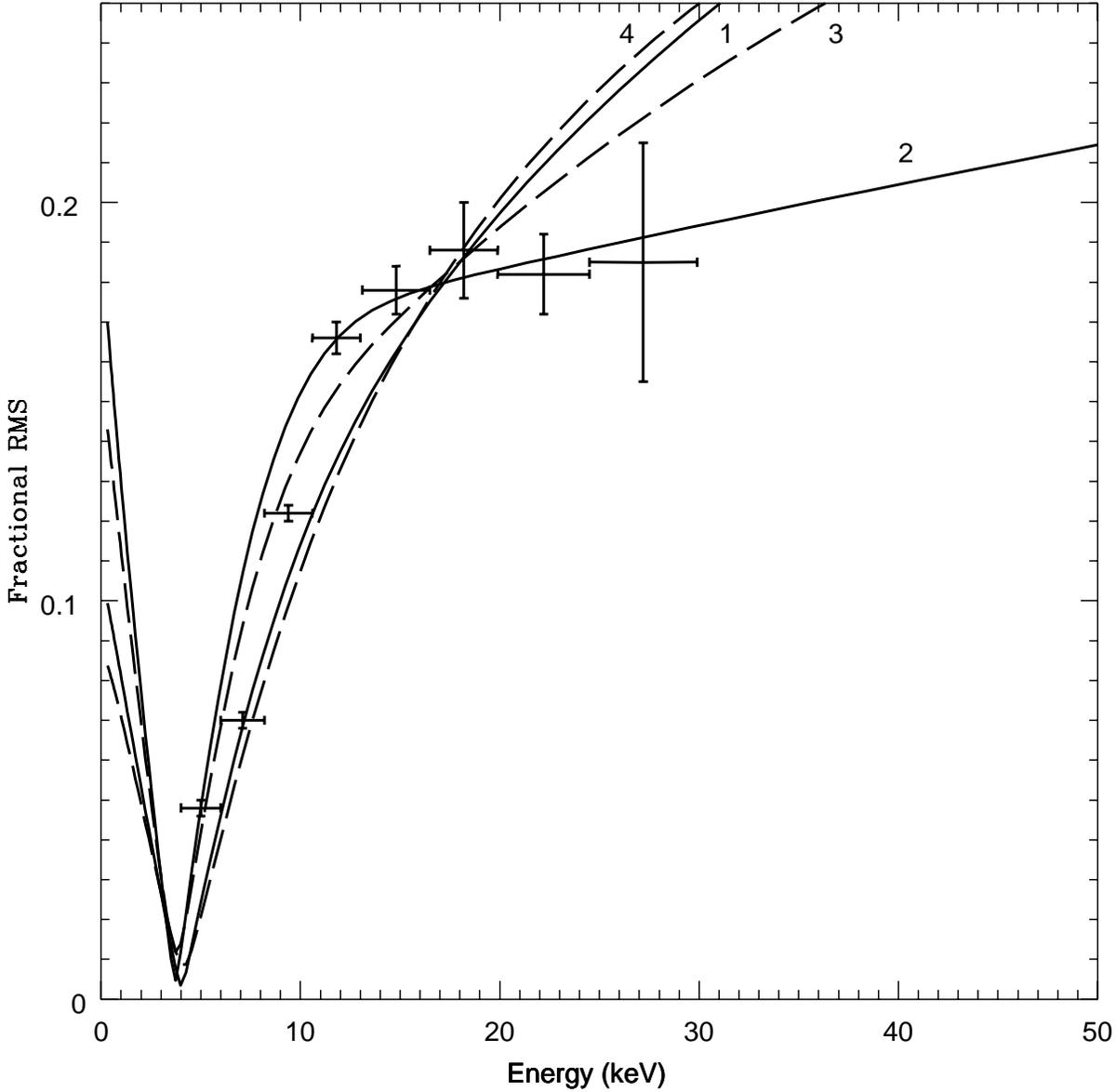}}}
\caption{Fractional RMS Amplitude vs. Energy.
The model predictions for $l=5$ km (solid curves) and $l=10$ km (dashed curves)
are compared to the data points from Berger et al.~1996.  The vertical error
bars denote one sigma errors, while the horizontal bars represent the energy
bands used but do not account for the uncertainty in the PCA's energy
resolution.  Note that model curves 1 and 3 overlap.
Curve 1: $f = 0.5$, $l = 5$ kms and
$\delta T_s = 0.58 \delta T_c exp(i 0.16 \pi)$.
Curve 2: $f = 0.75$, $l = 5$ kms and
$\delta T_s = 1.75 \delta T_c exp(i 0.46 \pi)$. 
Curve 3: $f = 0.5$, $l = 10$ kms and
$\delta T_s = 0.45 \delta T_c exp(i 0.27 \pi)$. 
Curve 4: $f = 0.75$, $l = 10$ kms and
$\delta T_s = 0.88 \delta T_c exp(i 0.48 \pi)$.  
.}
\end{figure*}

\begin{figure*}[h]
\hspace{-1.5cm}
{\mbox{\psboxto(17cm;20cm){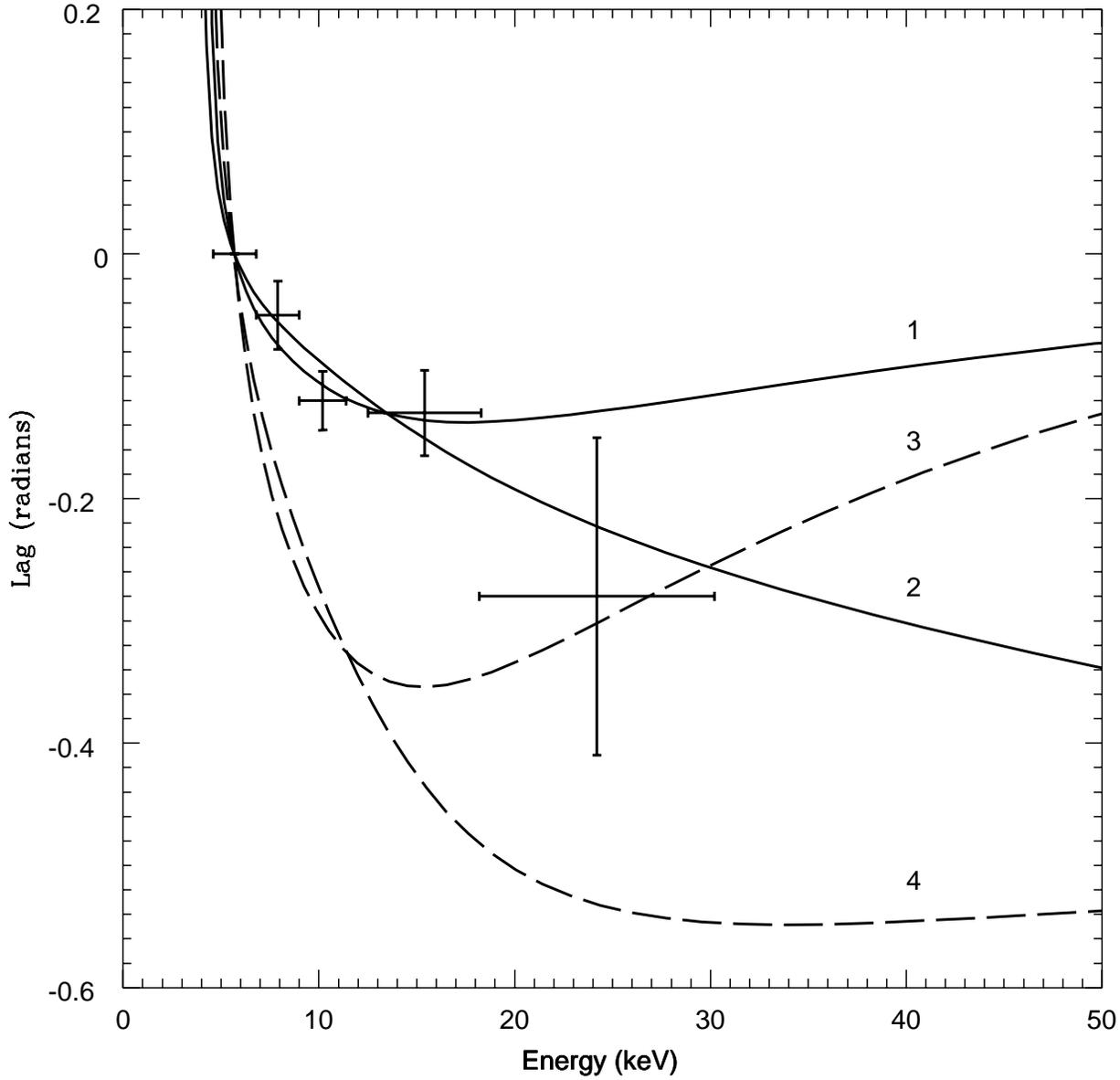}}}
\caption{Relative Phase Lag (Radians) vs. Energy.
The curves are model predictions for the same parameters
as in figure 1.  The data are taken from Vaughan et al 1998; as in Figure 1, the
vertical bars are one sigma errors and the horizontal bars show the energy
binning of the data.
.}
\end{figure*}

\end{document}